\documentclass[twocolumn,showkeys,showpacs,prb]{revtex4}
\usepackage{amsfonts, amsmath, latexsym}
\usepackage[dvips]{graphicx}

\newtheorem{3}{Theorem}

\begin{document}

\title{Moments of spectral functions: Monte Carlo evaluation and verification}

\author{Cristian Predescu}
\email{cpredescu@comcast.net} 
\affiliation{Department of Chemistry and Kenneth S. Pitzer Center for Theoretical Chemistry, University of California, Berkeley, California 94720}

\date{\today}
\begin{abstract}
The subject of the present study is the Monte Carlo path-integral evaluation of the  moments of spectral functions. Such moments can be computed by formal differentiation of certain estimating functionals that are infinitely-differentiable against time whenever the potential function is arbitrarily smooth. Here, I demonstrate that the numerical differentiation of the estimating functionals can be more successfully implemented by means of pseudospectral methods (e.g., exact differentiation of a Chebyshev polynomial interpolant), which utilize information from the entire interval $(-\beta \hbar / 2, \beta \hbar/2)$. The algorithmic detail that leads to robust numerical approximations is the fact that the path integral action and not the actual estimating functional are interpolated. Although the resulting approximation to the estimating functional is non-linear, the derivatives can be computed from it in a fast and stable way by contour integration in the complex plane, with the help of the Cauchy integral formula (e.g., by Lyness' method). An interesting aspect of the present development is that Hamburger's conditions for a finite sequence of numbers to be a moment sequence provide the necessary and sufficient criteria for the computed data to be compatible with the existence of an inversion algorithm. Finally, the issue of appearance of the sign problem in the computation of moments, albeit in a milder form than for other quantities, is addressed.
\end{abstract}
\maketitle

\section{Introduction}
Perhaps one of the most outstanding insufficiencies of the path integral formulation of quantum mechanics is that it does not lead directly to efficient algorithms for the computation of dynamical information. In contrast, statistical quantities or imaginary-time data are relatively easy to evaluate and the mechanisms to do so are well understood.\cite{Fey65, Sim79, Cep95} In principle, it is possible to relate the quantum correlation functions in real time to their imaginary-time counterparts by analytical continuation.\cite{Bay61, Jar96} However, such attempts lead to inverse problems  which, albeit uniquely determined, are fundamentally ill-posed. An example is represented by the real inverse Laplace transform,\cite{Jar96} a technique that has been extensively utilized as a link between the real and imaginary-time worlds. It requires the resolution of a linear integral equation that becomes extremely ill-conditioned upon discretization. The treatment of such problems is the domain of regularization theory, which attempts to stabilize the resulting equations by controlling certain properties of their solution.

Notwithstanding earlier attempts that were more or less in tone with Tikhonov's least-square approach,\cite{Sch85, Whi89, Jar89} the pioneering research of Gubernatis, Jarrell, Silver, and Sivia\cite{Gub91} has spearheaded the application of  methods of Bayesian statistical inference with entropic priors\cite{Gub91, Jar96} as a regularization technique for the inversion of the Laplace transform. To give a few examples, although somewhat restricted to the direct research interests of the present author, such methods have been successfully applied for the computation of various physical properties such as spectra,\cite{Gal94, Kim97, Kim98, Kri01} quantum rates of reaction,\cite{Rab00, Sim01} and diffusion constants.\cite{Rab02} When we say ``successfully,'' we take into account the fact that, in most cases, there are virtually no computationally feasible alternatives: the techniques based on imaginary-time data are amenable to direct Monte Carlo path integral treatment and exhibit little degradation of their stability with the increase in the physical dimensionality. This is so because the stability is related to the properties of the spectral function (a one-dimensional probability distribution) and not to the dimensionality of the physical system. 

Nevertheless, as Jarrell and Gubernatis point out,\cite{Jar96} ``to solve an ill-posed problem, nothing beats good data.'' The present paper does not attempt to improve on previous results regarding the stabilization of the inverse problems. Rather, its purpose is to provide a means to obtain high-quality input data for the reconstruction of various autocorrelation functions for physical systems in the continuum space. A recent study of the present author\cite{Pre04-70} suggests that one way to achieve better results is to break the inverse Laplace problem into two separate steps:  computation of moments by differencing an estimating functional (which is related to the imaginary-time correlation function) followed by  resolution of the ensuing symmetric Hamburger moment problem. Both steps are exponentially unstable, albeit to a lesser degree than the original problem. Very likely, their combined effect is a problem that is as ill-conditioned as the original one. Quite clearly, we cannot create new information in a stable and consistent manner just by manipulating the data. Is there any gain in the new approach? 

The reason we shy away from exponentially unstable algorithms is that they require the utilization of exponentially fast algorithms (polynomial in the number of digits) for the computation of the input data.  Unfortunately, most of the algorithms we posses converge polynomially in the best circumstances and there are only a few instances of exponentially fast algorithms. Some of the more interesting examples are related to the execution time of elementary functions on a classical computer, in arbitrary precision. In fact, Brent\cite{Bre76} has shown that the evaluation of elementary functions can be performed in time proportional to $O(n\log(n)^2\log\log(n))$ with respect to the number of digits $n$. Therefore, as Ref.~\onlinecite{Pre04-70} argues, differencing an arbitrarily smooth estimating functional  is a computationally feasible approach  whenever empirical potentials are utilized. The present author has obtained excellent results in unpublished tests that have employed Amber force fields\cite{Amb95} and Bailey's arbitrary precision package MPFUN90.\cite{Bai95}

However, such a technique cannot be applied for potentials that are the result of a computation performed in polynomial time. In addition, there is some unease related to the mere requirement of arbitrary precision. In most circumstances, we are interested in learning the properties of the spectral function in the low frequency region (for some transport phenomena, we are interested in the value at the origin of the spectral function associated with the flux-flux\cite{Rab00, Sim01} or velocity-velocity\cite{Rab02} correlation functions). Due to quantum and thermal smoothing, the low-frequency portion of the spectral function is largely insensitive to the precision with which the potential is known. Clearly, whether we use an otherwise smooth potential with $7$ (single precision) or $15$ (double precision) significant digits, we do not expect the value of a diffusion coefficient to change dramatically and this expectation is justified in many cases by results of perturbation theory. It follows that the instability of the differencing step is a property of the algorithm and not necessarily an inner characteristic of the problem.  

In Section~IV, we show that the instability associated with the differencing step can be removed by  interpolation of the action. That is, the path-integral action, regarded as a function of the imaginary time on the interval $(-\beta \hbar / 2, \beta \hbar / 2)$, is replaced by a smooth interpolant constructed by means of trigonometric or Chebyshev polynomials. For infinitely differentiable potentials, the interpolant converges faster than any polynomial and it rapidly feels the discontinuities due to the finite precision in the computation of the action. However, if it is known that the potential is smooth, one can utilize a low-degree interpolant only. By the arguments in the preceding section, the properties of the spectral functions in the low-frequency region and, therefore, the values of the low-order moments are not sensitive to the errors in the action. Despite the fact that the interpolated action may not necessarily come from a perturbed potential, the numerical results of Section~V suggest that the computed Monte Carlo data still represent a sequence of moments, even for low-order interpolants. 

We commence this paper, however, with Section~II, where we present a short review of the moment problem and discuss its relevance for the  reconstruction of spectral functions. In Section~III, benefitting from the existence of certain mathematical results concerning the Hamburger moment problem, we give necessary and sufficient criteria for inversion algorithms to exist. These criteria can be utilized as an a posteriori verification tool for the computed data and they reveal the exponential extent of the instability of the moment problem. 
Nevertheless, this instability has to be weighted against the fact that the information furnished even by a few tens of moments is, in general, quite substantial.  In Section~VI, we review the main findings of the present work and enumerate several research issues left outstanding.

\section{The inverse problem and the positivity requirement for the spectral function}

The dynamical quantity we want to evaluate is a quantum correlation function of the type\cite{Dol99r}
\begin{equation}
\label{eq:1}
C_{O,\lambda}(t) = \mathrm{tr}\left(e^{-(\beta/2+\lambda+it/\hbar) H}O^\dagger e^{-(\beta/2-\lambda -it/\hbar)H}O\right).
\end{equation}
The operator $H$ stands for the Hamiltonian of the system,  a self-adjoint and bounded from below operator, whereas $t \in \mathbb{R}$ and $\beta = 1/(k_B T) > 0$ are the real time and the inverse temperature, respectively. $O^\dagger$ denotes the adjoint of the operator $O$. The parameter $\lambda$ may take values in the interval $(-\beta/2, \beta/2)$. The values of the correlation functions for $\lambda = \pm\beta/2$ can be recovered as the corresponding limits, by continuity arguments. Many quantities of physical interest are related to the quantum correlation function defined by $\lambda = \beta/2$. However, the correlation functions defined by the parameter $\lambda$ are related one to each other by simple identities in the Fourier space.

Let us consider the associated spectral functions, which are defined by the Fourier transforms 
\begin{equation}
\label{eq:2}
\bar{C}_{O,\lambda}(\omega)=\frac{1}{2\pi} \int_{\mathbb{R}} e^{-i\omega t} C_{O,\lambda}(t) dt.
\end{equation}
With the help of the identity
\begin{equation}
\label{eq:3}
e^{-\beta_c H} = \int_{\mathbb{R}} e^{-\beta_c E} |E\rangle \langle E | dE,
\end{equation}
which is valid for all complex $\beta_c$ with $\mathrm{Re}(\beta_c) \geq 0$, 
 one computes
\begin{eqnarray*}
\bar{C}_{O,\lambda}(\omega) = \int_\mathbb{R} \int_\mathbb{R} e^{-(\beta/2-\lambda) E - (\beta/2+\lambda) E'}  |\langle E| O | E' \rangle|^2 \\ \times \left[\frac{1}{2\pi} \int_{\mathbb{R}} e^{it[-\omega + (E - E')/\hbar]} dt \right] dEdE'.
\end{eqnarray*}
Simple manipulations by means of the Fourier representation of the delta function show that
\begin{equation}
\label{eq:4} 
\bar{C}_{O,\lambda}(\omega) = \hbar e^{- (\beta/2-\lambda) \omega \hbar}\int_\mathbb{R}  e^{-\beta E } |\langle E+\omega \hbar| O | E \rangle|^2dE.
\end{equation}
Therefore, the Fourier transforms of the correlation functions are non-negative distributions (in fact, they are thermal averages of  certain power spectra). If we denote the special value for $\lambda = 0$ by $\bar{G}_{O}(\omega)$, then we have the relationship
\begin{equation}
\label{eq:5}
\bar{C}_{O,\lambda}(\omega) = e^{\lambda \omega \hbar} \bar{G}_{O}(\omega),
\end{equation}
which shows that all correlation functions defined by Eq.~(\ref{eq:1}) carry essentially the same information. It is not difficult to see that the spectral function $\bar{G}_{O}(\omega)$ is symmetrical about the origin and we shall refer to the corresponding correlation function as the thermally-symmetrized correlation function.

Berne and Harp\cite{Ber70} have pointed out that the computation of thermally-symmetrized quantum correlation functions 
\begin{equation}
\label{eq:6}
G_{O}(t)= \mathrm{tr}\left(e^{-\overline{\beta_c} H}O^\dagger e^{-\beta_c H }O\right),
\end{equation}
with $\beta_c = \beta /2 - it/\hbar$ and $\overline{\beta_c} = \beta /2 + it/\hbar$, might be an easier computational task. Certain quantities of physical interest, such as rates of reaction or diffusion constants, only depend on the value at the origin of the spectral functions, value that is independent of the particular choice of $\lambda$.  Miller, Schwartz, and Tromp\cite{Mil83} have utilized this independence to point out that the thermally-symmetrized flux-flux correlation function has better mathematical properties than the Yamamoto flux-flux correlation function,\cite{Yam60} which corresponds to an average over $\lambda$ on the interval $(-\beta/2, \beta/2)$. More recently, Predescu and Miller\cite{Pre05-109} have argued that the thermally-symmetrized spectral function is the one for which the moments (and, in general, any short-time information) are the most sensitive with respect to changes in the values of the spectral function near the origin. In other words, for this particular choice of correlation function, the values of the spectral function near the origin are expected to have the best continuity properties with respect to variations in the moments.

From Eq.~(\ref{eq:1}) and the inverse Fourier transform for Eq.~(\ref{eq:2}), it follows that 
\[
G_{O}(it) = C_{O,-t/\hbar}(0) = \int_{\mathbb{R}} \bar{C}_{O,-t/\hbar}(\omega)d\omega,
\]
an equality that holds provided that $t \in (-\beta\hbar/2, \beta\hbar/2)$. On the other hand, from Eq.~(\ref{eq:5}), we obtain the Laplace identity
\begin{equation}
\label{eq:7}
G_{O}(it) = \int_{\mathbb{R}}e^{-t \omega} \bar{G}_{O}(\omega)d\omega, 
\end{equation}
the inverse of which is the thermally-symmetrized spectral function. 

Having reached this point in our presentation, we pause and ask whether or not Eq.~(\ref{eq:7}) uniquely determines the spectral function. The answer is affirmative and follows from different arguments, all of which are based on the fact that the spectral function is \emph{positive}.  Thus, one could follow the path of Baym and Mermin\cite{Bay61} and use positivity to argue that the integrand of Eq.~(\ref{eq:7}) is absolutely integrable. In turn, this implies that the correlation function $G_{O}(t)$ must be analytic in the complex plane on the strip defined by $|\mathrm{Im}(t)| < \beta\hbar/2$. Standard results of complex analysis then show that $G_{O}(t)$ for real $t$ and, therefore, $\bar{G}_{O}(\omega)$ are uniquely determined by the values of $G_{O}(it)$ on the interval $(-\beta\hbar/2, \beta\hbar/2)$. The absolute integrability of $e^{-t \omega}\bar{G}_{O}(\omega)$ plays an important role in the proof of uniqueness and it should not be taken easy. Indeed, if the integral in Eq.~(\ref{eq:7}) is only required to converge in the Cauchy principal value sense, then there exist an infinity of solutions, of which only one is positive [non-positive examples are furnished by the Fourier transforms of Eq.~(\ref{eq:11})]. The issue is relevant because both absolute integrability and analyticity are constraints on the set of admissible solutions that are very difficult to implement on a computer. By comparison, enforcement of positivity is a more achievable goal.  

If we also use the a priori information that $G_{O}(it)$ and  $\bar{G}_{O}(\omega)$ are symmetric, the problem that must by solved in the context of the inverse Laplace transform method is: find the positive and symmetric distribution $\bar{G}_{O}(\omega)$ that satisfies the equation
\begin{equation}
\label{eq:8}
G_{O}(it) = \int_{\mathbb{R}}\cosh(\omega t) \bar{G}_{O}(\omega)d\omega, 
\end{equation}
for all $t \in [0, \beta \hbar/ 2)$. The input data for this problem is usually a finite sequence of values of the imaginary-time correlation function on an equally-spaced grid $\left\{t_{n,j} = j\beta \hbar /(2n): \; n \geq 1, \, 0 \leq j < n\right\}$. Upon discretization, the functional equation exhibits multiple solutions and  becomes determinate only upon the specification of an inversion algorithm. The main problem a computational physicist has to face is that the original functional equation is ill-posed in the sense of Hadamard. Although the problem has a unique solution, this solution lacks continuity with the input data for virtually any computationally reasonable topology. For example, there are sequences of functions $f_\epsilon (t)$ with $|f_\epsilon(t) - G_{O}(it)| / |G_{O}(it)| < \epsilon $ for all $t \in [0, \beta \hbar/2)$,  such that the problem
\begin{equation}
\label{eq:9}
f_\epsilon(t) = \int_{\mathbb{R}}\cosh(\omega t) \bar{G}_{O}(\omega)d\omega, 
\end{equation}
has no solutions for any $\epsilon$. Thus, just by mere control of the relative errors in the input data, we are not even guaranteed an inversion algorithm, much less a sequence of approximations to the spectral function that converges to $\bar{G}_{O}(\omega)$ as $\epsilon \to 0$. 

Another approach to proving the uniqueness of the solution of the inverse Laplace transform is via moments.\cite{Pre04-70} First, one utilizes the positivity of the spectral function to demonstrate that the Taylor series of the correlation function $G_{O}(t)$  about the origin has a convergence radius equal to or larger than $\beta \hbar / 2$. The sequence of even derivatives  of the imaginary-time correlation function at the origin reads
\begin{equation}
\label{eq:10}
D_{2k} = \left.\frac{d^{2k}G_O(it)}{dt^{2k}}\right|_{t = 0} = \int_{\mathbb{R}} \bar{G}_O(\omega)\omega^{2k}d\omega.
\end{equation} 
The odd moments are zero, by the symmetry of the spectral function. The ensuing symmetric Hamburger moment problem is then demonstrated to be uniquely determined, thus both proving the uniqueness of the reconstructed spectral function and suggesting an alternative computational approach. Unfortunately, the inverse moment problem also lacks continuity with the input data. Thus, just by controlling the relative errors for the moments, we are not guaranteed that an inversion algorithm exists. Moreover, the finite moment problem may also be undeterminate. If determinate, the finite moment problem is exponentially unstable. These stability issues will be addressed in the following section.

We conclude this section by emphasizing again the requirement of positivity  for the reconstructed spectral function. For any spectral function $\bar{G}_{O}(\omega)$, there are modifications that satisfy both the full moment problem given by Eq.~(\ref{eq:10}) and the Laplace equation given by Eq.~(\ref{eq:8}), precisely for the same input data as the physical spectral function.  Such modifications can be obtained by adding some integrable and infinitely differentiable function  that vanishes within the interval $(-\beta \hbar /2 , \beta \hbar / 2)$ to the correlation function and then taking the Fourier transform. A specific example of a function that satisfies both the full moment problem and the Laplace equation  is provided by the Fourier transform of
\begin{equation}
\label{eq:11}
G_{O}^{(\alpha)}(t) = G_{O}(t)  \left\{
\begin{array}{lll}
1&, & \mathrm{if} \; |t| \leq \beta \hbar / 2, \\
1+\exp\left[\frac{\alpha}{1-2|t|/(\hbar \beta)}\right]&, & \mathrm{otherwise},
\end{array}\right. 
\end{equation}
for any arbitrary and positive parameter $\alpha$. Such a Fourier transform always has a non-vanishing negative part. Of course, in agreement with Baym and Mermin's analyticity argument, the modification to the correlation function expressed by Eq.~(\ref{eq:11}) is not analytical.  Nevertheless, there is no effective procedure to enforce analyticity numerically and the positivity of the spectral function comes in handy. We stress that this positivity must be enforced to machine accuracy: for many modifications, the negative part of the modified spectral function appears in the high frequency region and can be a very small number, difficult to recognize on a plot, even if the correlation functions are ``obviously'' different. This is just another manifestation of the  instability of the inverse problems. 

\section{Stability of the inverse finite moment problem and verification of the moment data}

We begin this section with a short review of the Hamburger moment problem. All mathematical information contained in the present section can be found in standard references on the moment problem.\cite{Akh65} A sequence of numbers $D_0, D_1, \ldots $ is called a moment sequence if there exists a non-negative distribution, say $\bar{G}_O(\omega)$, such that
\begin{equation}
\label{eq:12}
D_k = \int_{\mathbb{R}} \bar{G}_O(\omega)\omega^k d\omega.
\end{equation} 
Quite clearly, not all sequences of numbers are moment sequences. For example, any moment of even order must be a non-negative number. Even more, by the non-negativity of the distribution $\bar{G}_O(\omega)$, we also have
\begin{equation}
\label{eq:13}
\sum_{j,k=1}^n D_{j+k}a_{j}a_{k} = \int_{\mathbb{R}} \bar{G}_O(\omega) \bigg(\sum_{j= 0}^na_j \omega^j\bigg)^2 d\omega \geq 0 
\end{equation}
for all sets of number $a_0, a_1, \ldots, a_n$. In matrix language, the last inequality is equivalent to the condition that the Gram matrices 
\begin{equation}
\label{eq:14}
\Delta'_n = \left(
\begin{array}{lllll}
D_0 & D_1 & D_2 & \cdots & D_n   \\
D_1 & D_2 & D_3 & \cdots & D_{n+1}\\
D_2 & D_3 & D_4 & \cdots & D_{n+2}\\
\vdots & \vdots & \vdots & \ddots & \vdots \\
D_n & D_{n+1}& D_{n+2}& \cdots & D_{2n}
\end{array}
\right)
\end{equation}
are positive semi-definite (that is, their lowest eigenvalues must be greater or equal to zero). A standard result from matrix analysis says that the Hermitian matrix $\Delta'_n$ is positive semi-definite if and only if all determinants $|\Delta'_k|$ with $0 \leq k \leq n$ are non-negative. Therefore, a necessary condition for a sequence of numbers to be a moment sequence is that the matrices $\Delta'_n$ are positive semi-definite for all $n \geq 0$ or that the determinants $|\Delta'_n|$ are non-negative for all $n \geq 0$. Hamburger has demonstrated that these conditions  are also sufficient for a sequence of numbers to be a moment sequence. In addition, he has shown that, given a finite sequence $D_0, D_1, \ldots, D_{2n}$, the positive semi-definiteness of $\Delta'_n$ is sufficient for the finite moment problem to have at least a solution (obviously, such a solution is rarely unique). If the quantities $D_1, D_3, \ldots, D_{2n-1}$ are zero, then there exists at least one symmetric solution.

As shown by Eq.~(\ref{eq:14}), the Gram matrices $\Delta'_n$ have a very special structure: the skew-diagonals are made up from identical elements. Such matrices, whether Gram or not, are called Hankel matrices. For the symmetric Hamburger moment problem, the skew-diagonals corresponding to moments of odd order are zero. The importance of the positive semi-definiteness of the matrices $\Delta'_n$ can be understood in the context of Theorem~3 of Ref.~\onlinecite{Pre04-70}, which is a standard convergence theorem in probability theory. Namely, assuming that we are given a collection of moment sequences $D_0^{(n)}, D_1^{(n)}, \ldots, D_{2n}^{(n)}$ (the low-rank terms of which are allowed to change with $n$ for generality) and letting $\bar{G}_O^{(n)}(\omega)$ and $G_O^{(n)}(t)$ denote the associated spectral and correlation functions, respectively, the convergence 
\[
\lim_{n \to \infty} D_k^{(n)} = D_k, \ \forall \; k \geq 0
\]
implies 
\[
\lim_{n \to \infty} G_O^{(n)}(t) = G_O(t), \ \forall \; t \in \mathbb{R}. 
\]
This result appears to contradict our previous assertion that the moment problem lacks continuity with the input data. To the contrary, the theorem provides a means of approximating the exact correlation function. The explanation is that Theorem~3 requires the input data $D_0^{(n)}, D_1^{(n)}, \ldots, D_{2n}^{(n)}$ to be finite moment sequences and it is this requirement that lacks continuity with the input data.  More exactly, for any $\epsilon > 0$, there exist a rank $n$ and numbers $D_0^{(n)}, D_1^{(n)}, \ldots, D_{2n}^{(n)} $ such that $|D_k^{(n)} -D_k| \leq \epsilon |D_k|,\; \forall \; 0 \leq k \leq 2n$, yet the new data are not a moment sequence (their Gram matrix is not positive semi-definite). Let us consider a particular case where $\epsilon = 0.01$. Thus, we know all the moments with $1\%$ relative accuracy. Is this enough  to be able to generate a good  approximation to the correlation function? The answer is no. As the results in the remainder of the present section show, it is very likely that the data we posses do not form a moment sequence, even for moderately large $n$. 

For our symmetric problem, the moments of odd order are zero and, therefore, their value is exactly known. In agreement with the hypothesis of Theorem~3, we require of any computational procedure to be able to provide the even-order moments with controlled relative error. By making this relative error small, we may assume that the even-order moments are positive. According to the discussion in the preceding paragraph, for the reconstruction algorithm to converge to the exact result in the limit that the relative error for the even-order moments converges to zero, it is sufficient that the inequalities 
\begin{equation}
\label{eq:15}
\left(\sum_{j,k=0}^{n} a_j D_{j+k} a_k\right) \left/ \left(\sum_{j = 0}^n D_{2j} a_j^2 \right) \right. \geq 0
\end{equation}
are satisfied for all numbers $a_0, a_1, \ldots, a_n$. By the positivity of the quantities $D_{2j}$, this condition is, of course, equivalent to the one provided by Eq.~(\ref{eq:13}). However, it also takes into account the fact that the relative errors of the moments $D_{2i}$ are controlled. By making the substitution $a_j = a'_j /\sqrt{D_{2j}}$ in Eq.~(\ref{eq:15}), we see that the above inequality is equivalent to the condition that the Hermitian matrices $\Delta_n$ of entries 
\[
(\Delta_n)_{j,k} = D_{j+k}\left/\sqrt{D_{2j}D_{2k}}\right., \quad 0 \leq j,k \leq n
\]
are positive semi-definite. We summarize the findings obtained so far in the present section  in the following theorem, which gives sufficient criteria for the existence of well-posed inversion algorithms.  
\begin{3}
For each $n \geq 1$, let $D_{0}^{(n)}, D_{2}^{(n)}, \ldots, D_{2n}^{(n)}$ be a finite sequence of positive  moment data. Let $D_{k}^{(n)} = 0$, for $k = 1, 3, \ldots, 2n-1$, and assume that the  Hermitian matrices $\Delta_n$ of entries
\begin{equation}
\label{eq:16}
(\Delta_n)_{j,k} = D_{j+k}^{(n)}\left/\left(D_{2j}^{(n)}D_{2k}^{(n)}\right)^{1/2}\right., \quad 0 \leq j,k \leq n
\end{equation}
are positive semi-definite. Then there exists at least one symmetric trial spectral function $\bar{G}_O^{(n)}(\omega)$ of even moments $D_{0}^{(n)}, D_{2}^{(n)}, \ldots, D_{2n}^{(n)}$. Moreover, with ${G}_O^{(n)}(t)$ denoting the associated trial autocorrelation function,  the convergence
\begin{equation}
\label{eq:17}
\lim_{n \to \infty} D_k^{(n)} = D_k, \ \forall \; k \geq 0
\end{equation}
implies 
\begin{equation}
\label{eq:18}
\lim_{n \to \infty} G_O^{(n)}(t) = G_O(t), \ \forall \; t \in \mathbb{R}. 
\end{equation}
\end{3}
The upper index $(n)$, which was needed in the formulation of the theorem, will be dropped from now on. We shall use the notation $D_{2k}$ for the moment data and understand that they are subject to both systematic and statistical errors.

 In view of the above theorem, it is quite unfortunate that the matrices $\Delta_n$ are ill-conditioned (although they are better behaved than the matrices $\Delta'_n$). In a research born out of frustration with the numerical instabilities of an otherwise reasonable algorithm, Tyrtyshnikov\cite{Tyr94} has demonstrated that the condition number (the ratio between the largest and the smallest eigenvalues) of any positive semi-definite Hankel matrix grows at least exponentially. As adapted to our problem, Tyrtyshnikov's result states that
\begin{equation}
\label{eq:19}
\kappa(\Delta'_n) \geq 3\cdot2^{n-5}.
\end{equation}
A tighter bound has been given more recently by Beckermann,\cite{Bek00} who has demonstrated that
\begin{equation}
\label{eq:20}
\kappa(\Delta'_n) \geq \gamma_0^{n-1}/(16n),
\end{equation}
for $n \geq 3$. The quantity $\gamma_0 \approx 3.210$ is related to the so-called Catalan series, but the exact value is not important for our purposes. Nevertheless, Beckermann has demonstrated that this is the best estimate for the minimal value of the exponential factor $\gamma_0$. Thus, there are positive semi-definite Hankel matrices for which the exponential growth is exactly $\gamma_0^n$. For most other applications, the exponential factor
\begin{equation}
\label{eq:21}
\gamma = \lim_{n \to \infty} \left[\kappa(\Delta'_n)\right]^{1/n}
\end{equation}
is larger than $\gamma_0 \approx 3.210$ and may itself increase to infinity.

\begin{figure}[!tbp]
     \includegraphics[angle=270,width=8.5cm,clip=t]{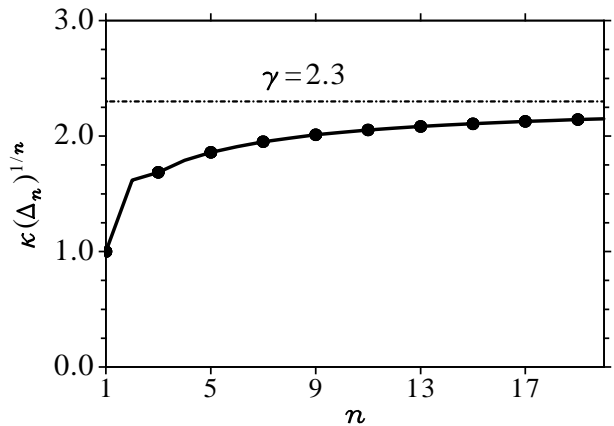}
   \caption[sqr] {\label{Fig:1} Asymptotic behavior of the quantities $\kappa(\Delta_n)^{1/n}$ for the flux-flux spectral function of a free particle. The asymptotic behavior demonstrates the exponential instability of the matrices $\Delta_n$. The condition number of the stability matrices increases roughly as $2.3^n$ for large $n$. }
\end{figure}

Although the matrices defined by Eq.~(\ref{eq:16}) are not Hankel, the extreme ill-conditioning of the matrices $\Delta'_n$ carries over to the matrices $\Delta_n$. A simple example will convince the reader of this. Consider the following spectral function\cite{Mil83,Mil98}
\begin{equation}
\label{eq:22}
\bar{G}_F(\omega) =\frac{1}{\beta h} \frac{|\omega| \hbar \beta}{2\pi} K_1\left(\frac{|\omega| \hbar \beta}{2}\right),
\end{equation}
where $K_1(x)$ denotes the respective modified Bessel function of the second kind. This is the spectral function associated with the thermally-symmetrized flux-flux correlation function (the so-called Miller, Schwartz, and Tromp correlation function,\cite{Mil83} or MST for short) for a free particle
\begin{equation}
\label{eq:23}
G_F(t) = \frac{1}{\beta h} \frac{(\beta \hbar / 2 )^2}{\left[t^2 + (\beta \hbar / 2 )^2\right]^{3/2}}.
\end{equation}
Formal differentiation of the correlation function at the origin shows that the even-order moments of the MST spectral function for a free particle are given by the equation
\begin{equation}
\label{eq:24}
D_{2k} = \frac{1}{\pi}\frac{2k+1}{(\beta \hbar)^{2k+2}}\frac{(2k)!^2}{k!^2}.
\end{equation} 
Using the above formula, one can set up the Hermitian matrices $\Delta_n$, diagonalize them, and compute their condition number $\kappa(\Delta_n)$. As apparent from Fig.~\ref{Fig:1}, the quantity $\kappa(\Delta_n)^{1/n}$ converges to a constant, the estimated value of which is $\gamma \approx 2.3$. This implies that the sequence of matrices $\Delta_n$ is exponentially unstable. As Eq.~(\ref{eq:4}) with $\lambda = 0$ suggests, the tail of the spectral function decays like $e^{-\beta \hbar \omega / 2}$ (as exponential order) for all thermally-symmetrized spectral functions. Because the relative values of the high-order moments depend only on the properties of the tail of the spectral function, we suggest that the  asymptotic value of $\gamma \approx 2.3$ may be characteristic of all spectral functions. The results computed for the symmetric Eckart barrier (they are presented in Fig.~\ref{Fig:2}) seem to support the suggestion. 

Why is the exponential instability of the matrices $\Delta_n$ so important? It shows that, just by controlling the relative errors of the even-order moments (that is, just by controlling the relative errors of the entries of the matrices $\Delta_n$), it is very likely that we cannot ensure the positive semi-definiteness of the matrices $\Delta_n$, even for moderately large $n$. As it is well known, the condition number of a matrix controls the relative errors in the values of the determinants $|\Delta_n|$ with respect to the relative errors in the entries of the matrix. Roughly speaking, $\log_{10}(\kappa(\Delta_n))$ represents the number of exact digits with which the entries of the matrix $\Delta_n$ must be known in order to guarantee that the determinant $|\Delta_n|$ is still positive. Therefore, the positive semi-definiteness of the matrices $\Delta_n$ must be a ``built in'' feature of the computational procedure.

\section{Moment estimating functionals}

As discussed in Ref.~\onlinecite{Pre04-70}, the moments can be computed by Monte Carlo integration as averages of some estimating functionals. A typical average is expressed by the equation
\begin{widetext}
\begin{eqnarray}\nonumber
\label{eq:25}&&
\frac{D_k}{\mathcal{N}_F} = \frac{1}{\mathcal{N}_F}\left. \frac{d^k}{dt^k}G_F(it)\right|_{t=0} \\ && =  \frac{\int_{\mathcal{S}}d \mathbf{x} d\mathbf{z}\mathbb{E}\mathbb{E}' e^{-\|\mathbf{z}\|^2}e^{-(\beta/2) \left[\int_0^1V\left(\mathbf{x} + \sigma_0 \mathbf{z} u + \sigma_0 B_u^0\right)du+\int_0^1V\left(\mathbf{x} + \sigma_0 \mathbf{z} u + \sigma_0 {B_u^0}'\right)du\right]} \frac{d^k}{dt^k}\mathcal{F}_t\left(\mathbf{x}, \mathbf{z}, B_\star^0, {B_\star^0}'\right)\Big|_{t = 0}}{\int_{\mathcal{S}}d \mathbf{x} d\mathbf{z}\mathbb{E}\mathbb{E}' e^{-\|\mathbf{z}\|^2} e^{-(\beta/2) \left[\int_0^1V\left(\mathbf{x} + \sigma_0 \mathbf{z} u + \sigma_0 B_u^0\right)du+\int_0^1V\left(\mathbf{x} + \sigma_0 \mathbf{z} u + \sigma_0 {B_u^0}'\right)du\right]} }.
\end{eqnarray}
\end{widetext}
The significance of the various terms can be found in the cited reference and will be partially explained below. Nonetheless, Eq.~(\ref{eq:25}) is sufficient to point out one of the main numerical difficulties of the present approach: due to the extreme complexity of the estimating functional $\mathcal{F}_t\left(\mathbf{x}, \mathbf{z}, B_\star^0, {B_\star^0}'\right)$, the differentiation against the parameter $t$ cannot be done analytically. As already mentioned in the introduction, the differentiation can be performed by finite difference. Such an approach, however, requires high precision evaluation of the potential function entering the expression of the estimating functional.

In the mathematical literature, it is well known that a superior technique for performing numerical differentiation is provided by the so-called pseudo-spectral methods. The functional $\mathcal{F}_t\left(\mathbf{x}, \mathbf{z}, B_\star^0, {B_\star^0}'\right)$ is infinitely differentiable on the interval $t \in (-\beta \hbar / 2, \beta \hbar / 2)$ provided that the potential function is also infinitely differentiable. The functional can  be approximated on some compact interval  $[-\theta\beta \hbar / 2, \theta\beta \hbar / 2]$, $0 < \theta < 1$, by Chebyshev polynomial interpolation (or, following certain transformations, one can also perform a trigonometric interpolation). The resulting Chebyshev polynomial can then be differentiated analytically. It can be demonstrated that, when differentiating such an interpolating polynomial, the error committed decays to zero exponentially fast for functionals that are analytical in $t$ (that is, for analytical potentials) and faster than any polynomial for infinitely differentiable functionals (potentials). This should be compared with the polynomial decay obtained by finite difference. A short exposition of these results and some elegant proofs can be found in Ref.~\onlinecite{Tad86}. 

The choice of the parameter $\theta$ is crucial in obtaining accurate estimates of the derivatives. If $\theta$ is too small, then the input information represented by the values of the estimating functional $\mathcal{F}_t\left(\mathbf{x}, \mathbf{z}, B_\star^0, {B_\star^0}'\right)$ for the Chebyshev knots in the interval $[-\theta\beta \hbar / 2, \theta\beta \hbar / 2]$ are highly redundant and the necessary precision is very high. Therefore, the parameter $\theta$ should be chosen large, as close as possible to the value $1$ (say $\theta = 3/4$). Even so, the number of Chebyshev coefficients that are accurately computed is not very great and depends on the precision of the machine. As a rule of thumb, one can rely upon $8$ to $16$ coefficients in single precision and $16$ to $32$ coefficients in double precision. As recommended in Numerical Recipes,\cite{Pre92} a technique that improves the quality of the polynomial interpolation is to truncate a higher-order interpolating polynomial to a lesser degree. We shall refer to such an interpolation polynomial as a regressed polynomial. 

Given the limitation due to the finite precision with which the potential can be evaluated, the pseudo-spectral technique may fail to provide adequate estimates for the derivatives if the estimating functional and its derivatives are not easily approximated by a low-degree interpolant. Unfortunately, this is the case for the present computational task. The culprit is the Boltzmann factor that enters the definition of the estimating functional. To understand how this factor enters our equations, we review the definition of the flux-flux estimating functional. \cite{Pre04-70} The surface through which the flux is computed is assumed to be a plane of equation $\mathbf{x}_1 = 0$. Thus, the space $\mathcal{S}$ appearing in Eq.~(\ref{eq:25}) is the subspace of $\mathbb{R}^d \times \mathbb{R}^d$ of equation $\mathbf{x}_1 = 0$ and $\mathbf{z}_1 = 0$. 
The quantities $B_u^0$ and ${B_u^0}'$ are independent $d$-dimensional standard Brownian bridges ($d$-dimensional vector valued stochastic processes, the components of which are independent one-dimensional standard Brownian bridges). We also define the entities $\beta_t$, $\sigma_t$, and $\sigma_{\pm t}$ as $\beta_t = \beta / 2 + t / \hbar$, $\sigma_t = \left(\hbar^2\beta_t / m_0\right)^{1/2}$, and $\sigma_{\pm t} = \sigma_t \sigma_{-t} / \sigma_0$, respectively. Finally, we let $V(x)$, $V'(x)$, and $V''(x)$ represent the potential and its first and second order partial derivatives against the reaction coordinate $\mathbf{x}_1$. 

The following notation, additional to what has been introduced in Ref.~\onlinecite{Pre04-70}, defines several action-like variables which will constitute the basic entities that we interpolate:
\begin{equation}
\label{eq:26}
S_t(\mathbf{x}, \mathbf{z}, B_*^0) = \beta_t \int_0^1V\left(\mathbf{x} + \sigma_{\pm t} \mathbf{z} u + \sigma_{t} B_u^0\right)du,
\end{equation}
\begin{equation}
\label{eq:27}
S'^{,a}_t(\mathbf{x}, \mathbf{z}, B_*^0) = \beta_t \int_0^1V'\left(\mathbf{x} + \sigma_{\pm t} \mathbf{z} u + \sigma_{t} B_u^0\right)udu,
\end{equation}
\begin{equation}
\label{eq:28}
S'^{,b}_t(\mathbf{x}, \mathbf{z}, B_*^0) = \beta_t \int_0^1V'\left(\mathbf{x} + \sigma_{\pm t} \mathbf{z} u + \sigma_{t} B_u^0\right)(1-u)du,
\end{equation}
and
\begin{equation}
\label{eq:29}
S''_t(\mathbf{x}, \mathbf{z}, B_*^0) = \beta_t \int_0^1V''\left(\mathbf{x} + \sigma_{\pm t} \mathbf{z} u + \sigma_{t} B_u^0\right)u(1-u)du.
\end{equation}
In terms of the action-like variables, we define
\begin{eqnarray}
\label{eq:30}
\nonumber
\mathcal{F}_t^0\left(\mathbf{x}, \mathbf{z}, B_\star^0, {B_\star^0}'\right) &=&  \frac{1}{\sigma_{-t}^2} + \frac{1}{ \sigma_{t}^2} \\ \nonumber &+& \left[S'^{,a}_t(\mathbf{x}, \mathbf{z}, {B_*^0}')-S'^{,a}_{-t}(\mathbf{x}, \mathbf{z}, B_*^0)\right]\\ &\times& \left[S'^{,b}_t(\mathbf{x}, \mathbf{z}, {B_*^0}') -S'^{,b}_{-t}(\mathbf{x}, \mathbf{z}, B_*^0)\right] \qquad \\ \nonumber
&-&S''_t(\mathbf{x}, \mathbf{z}, {B_*^0}')- S''_{-t}(\mathbf{x}, \mathbf{z}, B_*^0) 
\end{eqnarray}
and
\begin{eqnarray}
\label{eq:31} 
 \Delta_{t}\left(\mathbf{x}, \mathbf{z}, B_\star^0\right) = \frac{2}{\beta}\left[ S_0(\mathbf{x}, \mathbf{z}, B_*^0) - S_t(\mathbf{x}, \mathbf{z}, B_*^0)\right].
\end{eqnarray}
Then, as shown in Ref.~\onlinecite{Pre04-70}, 
\begin{eqnarray}
\label{eq:32}
 \nonumber 
\mathcal{F}_t\left(\mathbf{x}, \mathbf{z}, B_\star^0, {B_\star^0}'\right)= \frac{1}{2\sqrt{\beta_{-t}\beta_t}}   \left\{\mathcal{F}_{-t}^0\left(\mathbf{x}, \mathbf{z}, B_\star^0, {B_\star^0}'\right) \right. \\ \times  e^{(\beta/2)\left[ \Delta_{t}\left(\mathbf{x}, \mathbf{z}, B_\star^0\right) + \Delta_{-t}\left(\mathbf{x}, \mathbf{z}, {B_\star^0}'\right)\right]} + \mathcal{F}_t^0\left(\mathbf{x}, \mathbf{z}, B_\star^0, {B_\star^0}'\right) \quad \\ \left. \times   e^{(\beta/2)\left[ \Delta_{-t}\left(\mathbf{x}, \mathbf{z}, B_\star^0\right) + \Delta_t\left(\mathbf{x},\mathbf{z},{B_\star^0}'\right)\right]}\right\}. \nonumber
\end{eqnarray}

The weighting factors of the type
\[
e^{(\beta/2)\left[ \Delta_{t}\left(\mathbf{x}, \mathbf{z}, B_\star^0\right) + \Delta_{-t}\left(\mathbf{x}, \mathbf{z}, {B_\star^0}'\right)\right]}
\] 
may vary quite violently for low enough temperatures and cannot be readily approximated with a low-degree interpolant even for smooth potentials. On the other hand, the action-like quantities do not vary violently with $t$ even for low temperatures. The extent to which they vary is controlled by the values of the potential or its derivatives (the smoothness of these functions is assumed). It is, of course, these quantities that we intend to approximate by pseudo-spectral methods.

For this purpose and in order to take advantage of the whole array of values $t \in [-\beta \hbar / 2, \beta \hbar / 2]$, we make the substitution
\begin{equation}
\label{eq:33}
\sqrt{\frac{1}{2} + \frac{t}{\beta \hbar}} = \sin\left[\frac{\pi}{4}(1 + \phi)\right]
\end{equation}
and regard the action-like entities as functions of $\phi$ on the interval $[-1, 1]$. With the notation $\sigma = (\hbar^2\beta/m_0)^{1/2}$, the equalities 
\begin{eqnarray}
\label{eq:34}\nonumber
\beta_t & = & \beta \sin\left[\frac{\pi}{4}(1 + \phi)\right]^2,\\
\sigma_t & = & \sigma\sin\left[\frac{\pi}{4}(1 + \phi)\right],\\
\sigma_{\pm t} & = & \frac{1}{2}  \sigma\sin\left[\frac{\pi}{2}(1 + \phi)\right] \nonumber
\end{eqnarray}
show, for example, that the action $S_t$ can be regarded as the function of $\phi$ given by the formula
\begin{eqnarray*}
\tilde{S}_\phi(\mathbf{x}, \mathbf{z}, B_*^0) = 
\beta \sin\left[\frac{\pi}{4}(1 + \phi)\right]^2  \int_0^1V\Big(\mathbf{x} + \frac{1}{2}  \sigma \mathbf{z} u \\ \times \sin\left[\frac{\pi}{2}(1 + \phi)\right]   + \sigma B_u^0\sin\left[\frac{\pi}{4}(1 + \phi)\right]\Big)du.
\end{eqnarray*}
This function is infinitely differentiable on the interval $[-1, 1]$ whenever the potential is so. Therefore, the action-like functionals can be Chebyshev approximated on this interval faster than any polynomial. An alternative interpolation procedure utilizes trigonometric polynomials and is based on the observation that the action-like variables are periodical in $\phi$. Unfortunately, the period is $8$ and the functionals must be sampled on the larger interval $\phi \in [-4,4]$, which is more costly, because one needs roughly $4$ times more points. Although the trigonometric (Fourier) interpolation has the nice property that it becomes exact for potentials that are polynomials, we did not notice any significant advantage over the Chebyshev interpolation in more realistic numerical tests (the two techniques behave in a similar fashion for comparable meshes of the interpolatory knots). 

For some order of approximation $n$, let $\phi_k = \cos(\pi(k -0.5)/n)$, $k = 1, 2, \ldots, n$, be the nodes of the Chebyshev polynomial $T_n(x)$. For $j = 0, 1, \ldots, n-1$, the Chebyshev coefficients are given by 
\begin{equation}
\label{eq:35}
c_j(\mathbf{x}, \mathbf{z}, B_*^0) = \frac{2}{n}\sum_{k = 1}^n \tilde{S}_{\phi_k}(\mathbf{x}, \mathbf{z}, B_*^0) \cos\left[\frac{\pi j (k - 0.5)}{n}\right],
\end{equation} 
and we have the following approximation formula
\begin{equation}
\label{eq:36}
\tilde{S}_{\phi}(\mathbf{x}, \mathbf{z}, B_*^0) \approx \frac{1}{2}c_0(\mathbf{x}, \mathbf{z}, B_*^0) +\sum_{k = 1}^{n-1} c_k(\mathbf{x}, \mathbf{z}, B_*^0) T_k(\phi).
\end{equation} 
The right-hand side expression in Eq.~(\ref{eq:36}) is called the Chebyshev interpolation polynomial of rank $n - 1$. This polynomial is the unique polynomial of rank $n-1$ that coincides with $\tilde{S}_{\phi}(\mathbf{x}, \mathbf{z}, B_*^0)$ for all $n$ interpolatory knots $\phi_k$. 

Of course, interpolation polynomials must be computed for all action-like functionals described by Eqs.~(\ref{eq:26}) to (\ref{eq:29}). The computation of the Chebyshev coefficients can be performed in a fast and stable way by cosine fast Fourier transform (FFT).\cite{Pre92} We recommend the use of such a transform not so much for reasons of computational efficiency as for reasons of accuracy. If $\epsilon$ is the floating-point relative precision of the machine, the relative error for the Cooley-Tukey FFT algorithm is $O(\epsilon \log(n))$, compared to $O(\epsilon n^{3/2})$ for the direct matrix multiplication technique.\cite{Gen66} The numerical evaluation of the action-like functionals from their Chebyshev coefficients is to be performed by the Clenshaw recurrence formula, which is documented in Numerical Recipes.\cite{Pre92}

Replacing the action-like functionals with their Chebyshev approximation in Eqs.~(\ref{eq:30}) to (\ref{eq:32}), we obtain a non-linear approximation  $\tilde{\mathcal{F}}_\phi\left(\mathbf{x}, \mathbf{z}, B_\star^0, {B_\star^0}'\right)$ of the estimating functional in terms of the variable $\phi$. In terms of the variable $t$, one has
\begin{equation}
\label{eq:37}
\mathcal{F}_t\left(\mathbf{x}, \mathbf{z}, B_\star^0, {B_\star^0}'\right) \approx \tilde{\mathcal{F}}_{\phi(t)}\left(\mathbf{x}, \mathbf{z}, B_\star^0, {B_\star^0}'\right),
\end{equation}
where 
\begin{eqnarray}
\label{eq:38} \nonumber
\phi(t) &=& -1 + \frac{4}{\pi}\arcsin\left(\sqrt{\frac{1}{2}+ \frac{t}{\hbar \beta}}\right) \\ &=& -1 - \frac{4i}{\pi}\ln\left(\sqrt{\frac{1}{2}- \frac{t}{\hbar \beta}}+i\sqrt{\frac{1}{2}+ \frac{t}{\hbar \beta}}\right)
\end{eqnarray} 
is, of course, the appropriate solution of  Eq.~(\ref{eq:33}). Eq.~(\ref{eq:38}), which involves the use of complex numbers, already betrays the approach we shall utilize to compute the derivatives of the estimating function at the origin: contour integration of the complex extension of the non-linear Chebyshev approximation in the complex plane. The numerical algorithm utilized is due to Lyness\cite{Lyn68} and is summarized in the following paragraph. 

The Cauchy integral formula
\begin{eqnarray}
\label{eq:39} \nonumber &&
\frac{d^k}{dt^k}\mathcal{F}_t\left(\mathbf{x}, \mathbf{z}, B_\star^0, {B_\star^0}'\right)\Big|_{t = 0} \\ &&= \frac{k!}{2\pi i}\int_{C}\frac{1}{t^{k+1}}\mathcal{F}_t\left(\mathbf{x}, \mathbf{z}, B_\star^0, {B_\star^0}'\right)dt
\end{eqnarray} 
provides a way to compute the derivatives of an analytical function by computing integrals. The contour $C$ must be a closed curve that contains the origin in its interior. It will be taken to be a circle of radius $r\in(0, \beta \hbar/2)$ centered about the origin.  Eq.~(\ref{eq:39}) becomes
\begin{equation}
\label{eq:40} 
\frac{k!}{r^k}\int_{0}^1e^{-2\pi i k \theta}\mathcal{F}_{r e^{2\pi i \theta}}\left(\mathbf{x}, \mathbf{z}, B_\star^0, {B_\star^0}'\right)d\theta
\end{equation} 
and shows that the problem of computing derivatives is equivalent with that of evaluating the Fourier coefficients of a periodic function. The one-dimensional integral in Eq.~(\ref{eq:40}) is to be computed by trapezoidal quadrature. We have the approximation
\begin{eqnarray}
\label{eq:41} \nonumber &&
\frac{d^k}{dt^k}\mathcal{F}_t\left(\mathbf{x}, \mathbf{z}, B_\star^0, {B_\star^0}'\right)\Big|_{t = 0}\\ && \approx 
\frac{k!}{r^k}\frac{1}{m}\sum_{j = 1}^{m}e^{-2\pi i k j / m}\mathcal{F}_{r e^{2\pi i j/m}}\left(\mathbf{x}, \mathbf{z}, B_\star^0, {B_\star^0}'\right), \qquad
\end{eqnarray} 
where the summation is best executed by FFT. Lyness\cite{Lyn68} has demonstrated that this simple-looking algorithm is numerically stable and converges exponentially fast with respect to the number of quadrature knots $m$. For this reason, the algorithm is known as Lyness' method. Although criteria for choosing optimal values for the radius $r$ are known, numerical tests show that the ad-hoc value of $r = \beta \hbar / 4$ produces excellent results. 

We conclude the present section by pointing out an issue of convergence and precision. Unless the potential function is analytical, one cannot extend the estimating functional $\mathcal{F}_t\left(\mathbf{x}, \mathbf{z}, B_\star^0, {B_\star^0}'\right)$ to the entire complex plane. Thus, the non-linear Chebyshev approximation, although convergent on the real interval $(-\beta \hbar / 2, \beta \hbar / 2)$, diverges on the complex disk of radius $\beta \hbar / 2$, as the rank $n$ of the interpolating polynomial goes to infinity. We take advantage of the rapid convergence and the stability of Lyness' method to compensate for this divergence. However, to also compensate for the loss in precision in the evaluation of the integrand,  one may need to evaluate the Chebyshev polynomials as well as the final non-linear approximation in higher precision. Numerical tests show that this divergence is very weak and that the need for higher precision seems only theoretical. With the increase in the rank of the Chebyshev approximation beyond a certain level, we do notice a sudden divergence, but this  is caused by the limited precision with which the high-order coefficients are determined and appears even for analytical potentials. The convergence of the Chebyshev approximation is normally so fast that it rapidly feels  the lack of smoothness of the action due to the finite precision of the machine. Thus, the high-order coefficients, instead of decaying steadily to zero, remain roughly constant as magnitude. One counteracts this artifact by truncating the Chebyshev series to a safe number of coefficients (usually less than $16$ in single precision and less than $32$ in double precision).  The sudden divergence can be easily spotted by comparing the values  of $D_0$ computed with two slightly different estimators: the first one utilizes the estimating functional $\mathcal{F}_0\left(\mathbf{x}, \mathbf{z}, B_\star^0, {B_\star^0}'\right)$ (more precisely, to account for the case when the polynomials are regressed, the value at the origin of its non-linear approximation in terms of Chebyshev polynomials), whereas the  second one utilizes the Cauchy contour integral
\[
\frac{1}{m}\sum_{j = 1}^{m}\mathcal{F}_{r e^{2\pi i j/m}}\left(\mathbf{x}, \mathbf{z}, B_\star^0, {B_\star^0}'\right).
\]
The agreement for the computed values of $D_0$ must be better than the minimal accuracy that would guarantee the positive semi-definiteness of the stability matrices. 

\section{A numerical example: the symmetric Eckart barrier}

In order to demonstrate the capabilities of the present technique, we compute the first $20$ even-order moments of the flux-flux spectral function for a symmetric Eckart barrier at the low temperature of $T = 100~\mathrm{K}$. The parameters for the Eckart barrier are those also utilized in Ref.~\onlinecite{Pre04-70}. The potential is
\begin{equation}
\label{eq:50a}
V(x) = V_0 \; \mathrm{sech}(ax)^2,
\end{equation} 
with  $V_0 = 0.425~\mathrm{eV}$, $a = 1.36~\mathrm{a.u.}$, and $m_0 =  1060~\mathrm{a.u.}$ The parameters for the barrier  are chosen to correspond approximately to the $\mathrm{H}+\mathrm{H}_2$ reaction. As discussed in Section~III, the necessary and sufficient criteria that guarantee the existence of a convergent reconstruction algorithm are a) positivity and control of the relative errors for the moments and b) positive semi-definiteness of the matrices $\Delta_n$. 

Let us summarily describe the main features of the Monte Carlo path integral technique utilized. According to Eq.~(\ref{eq:25}), what we actually compute are the ratios $D_{2k}/\mathcal{N}_F$, where  
\begin{widetext}
\begin{equation}
\label{eq:43}
\mathcal{N}_F = \frac{1}{8\pi m_{0}}\left(\frac{1}{2\pi \sigma_{0}}\right)^{d-1}\int_{\mathcal{S}}d \mathbf{x} d\mathbf{z}\mathbb{E}\mathbb{E}'e^{-\|\mathbf{z}\|^2} e^{-(\beta/2) \left[\int_0^1V\left(\mathbf{x} + \sigma_0 \mathbf{z} u + \sigma_0 B_u^0\right)du+\int_0^1V\left(\mathbf{x} + \sigma_0 \mathbf{z} u + \sigma_0 {B_u^0}'\right)du\right]}.
\end{equation}
\end{widetext}
Although the normalization coefficient $\mathcal{N}_F$ can also be determined by Monte Carlo integration, its value is irrelevant for the present study. The path integral technique utilized is based on the fourth-order short-time approximation introduced in Ref.~\onlinecite{Pre04-69}. A Trotter index of $16$ has been employed, for a total of $64$ path variables. In  order to diminish the amount of correlation in the Monte Carlo chain, we have utilized the so-called fast sampling algorithm considered in Ref.~\onlinecite{Pre05-71} as well as the parallel tempering technique. The parallel tempering exchanges have been performed with replicas of temperatures arranged in geometric progression up to $5000~\mathrm{K}$. A total of $10$ million complete sweeps through the space of path variables have been made. These sweeps have been divided in $100$ blocks.

Naturally, the Monte Carlo simulation consists of two parts: sampling and accumulation of averages for the estimating functionals. Due to the nature of the fast sampling algorithm, which organizes the path variables in $2 + \log_2(64) = 8$ layers with the variables from the same layer sampled separately (and independently), the computational cost for a sweep is $64\cdot 8$ calls to the potential function. The computational cost for the estimating functionals is $64 \cdot 3 \cdot 32$ calls. The factor $3$ comes from the three different types of functions that are called [$V(\mathbf{x})$, $V'(\mathbf{x})$, and $V''(\mathbf{x})$] whereas the factor $32$ represents the number of Chebyshev knots. We mention that even for such a simple analytical potential, the largest part of the computation is spent on the evaluation of the potential and its derivatives and not on performing the numerical manipulations considered in the preceding section. In order to dedicate comparable amounts of time to the sampling and estimator evaluation parts, we have evaluated the estimators every $3 \cdot 32 / 8 = 12$ sweeps. We made this analysis just in order to warn the reader about the magnitude of the computational cost incurred by the evaluation of the estimating functionals for derivatives. Thus, it is not worth to accumulate averages after each sweep, especially given the usually large correlation of the Monte Carlo sampler (it is very rare that the correlation times are smaller than $12$ in realistic simulations). 

As mentioned before, the main purpose of the present development is to provide a technique that is capable of producing accurate estimates for the high order derivatives without the need to evaluate the potential function in arbitrary precision. Therefore, the potential function and its derivatives utilized in the construction of the estimating functionals have been evaluated in single precision (about $7$ significant decimal digits) only. This is also the precision with which the Chebyshev coefficients are evaluated.  On the other hand, the evaluation of the Chebyshev polynomials, the contour integration in the complex plane, and the accumulation of averages have been performed in quadruple precision (about $31$ significant decimal digits). The sampling part of the simulation has been conducted in double precision (about $15$ significant decimal digits). To complete the description of the algorithm, we mention that the Chebyshev polynomials have been regressed to the first $16$ coefficients and the contour integration has been performed in $64$ quadrature knots. 

We accumulate the averages in quadruple precision in anticipation of the loss of precision due to the exponential instability of the matrices $\Delta_n$. We stress that, although the moments $D_{2k}$ are determined with a precision of a few digits only, an important amount of information resides in the remaining imprecise digits. 
To understand how this may happen, let us analyze the problem of computing the moments if the spectral function is analytically known and can itself be sampled. Thus, we compute averages of the type
\begin{equation}
\label{eq:44}
\mu_{k} = D_{k}/D_0 = \int_{\mathbb{R}}\bar{G}_F(\omega)\omega^{k}d\omega \left/ \int_{\mathbb{R}}\bar{G}_F(\omega)d\omega  \right. 
\end{equation}
for $k = 0, 1, \ldots$ by Monte Carlo integration.  The actual ratios in $N$ sample points are 
\begin{equation}
\label{eq:45}
\mu_{k}^{(N)} = \frac{1}{N} \sum_{j = 1}^N \omega_j^{k} = \int_{\mathbb{R}}\rho_N(\omega)\omega^{k}d\omega,
\end{equation}
where 
\begin{equation}
\label{eq:46}
\rho_N(\omega) = \frac{1}{N} \sum_{j = 1}^N \delta(\omega - \omega_j)
\end{equation}
is some discrete measure that depends upon the history of the Monte Carlo integration. We notice that \emph{irrespective} of what this discrete measure is, the sequence of numbers $\{\mu_{k}^{(N)}; k = 0, 1, \ldots\}$ is a \emph{moment sequence} because it comes from some positive measure. It therefore satisfies the requirements of positive semi-definiteness of the stability matrices $\Delta_n$ regardless of the actual number of samples $N$. It also exhibits the same instability issues as the original problem. Theoretically, if we determine the whole sequence of moments $\{\mu_k^{(N)}; k = 0, 1, \ldots\}$ for some fixed $N$ with arbitrary precision, we can reconstruct the measure that has generated the sequence, that is, the distribution $\rho_N(\omega)$. But if these statistically inexact moments are not known with sufficient precision, the exponential instability will cause them to loose the crucial property that they form a moment sequence. 

The lesson we learn from the preceding exposition is that we cannot compute moments of different orders in independent Monte Carlo runs, because it is difficult to attain the  precision necessary to ensure the positive semi-definiteness of the stability matrices $\Delta_n$.   In this paper, we declare ourselves content with the simultaneous computation of the moments in the same Monte Carlo simulation. We mention, however, that this is not necessarily without penalty. If, for example, one is interested in evaluating the tail of the spectral function [which decays like $e^{-\beta \hbar \omega / 2}$, as Eq.~(\ref{eq:4}) with $\lambda = 0$ suggests], then there are going to be problems related to the fact that the tail of the distribution is infrequently sampled by the Monte Carlo walker because of its exponentially vanishing statistical weight. Accordingly, the information contained in the moments determined by Monte Carlo integration poorly reproduces the properties of the tail. Thus, our assumption is that we are interested in the properties of the spectral function near the origin or in regions of important statistical weight.

\squeezetable
\begin{table*}[!htbp]
\caption{\label{Tab:I} Moments of even order and their percent relative statistical errors determined by Monte Carlo integration. A number of $\log_{10}[\kappa(\Delta_{20})] \approx 8$ significant figures are necessary in order to prevent the loss of the positive semi-definiteness property of the stability matrices. We give the results with one significant digit more than the minimal requirement so that the reader may verify that the matrix $\Delta_{20}$ is positive definite.}
\begin{tabular}{| c | c | c | c | c | c | c | c |} 
\hline 
$k$ & $0$ & $1$ & $2$ & $3$ & $4$ & $5$ & $6$ \\
\hline
$D_{2k}/\mathcal{N}_F$ & $5.43598845\cdot 10^1$ & $2.25127499\cdot 10^{-4}$ & $3.79457212\cdot 10^{-9}$ & $1.31959326\cdot 10^{-13}$ & $7.49921061\cdot 10^{-18}$ & $6.09205035\cdot 10^{-22} $ & $6.48529540\cdot 10^{-26}$\\
\hline
Error & $0.5\%$ & $0.7\%$ & $1.2\%$ & $2.4\%$ & $4.3\%$ & $6.7\%$ & $9.5\%$  \\
\hline \hline
$k$ & $7$ & $8$ & $9$ & $10$ & $11$ & $12$ & $13$ \\
\hline
$D_{2k}/\mathcal{N}_F$ & $8.53623493\cdot 10^{-30}$ & $1.33546663\cdot 10^{-33}$ & $2.41660791\cdot 10^{-37}$ & $4.96421869\cdot 10^{-41}$ & $1.14347963\cdot 10^{-44}$ & $2.93267228\cdot 10^{-48} $ & $8.35072926\cdot 10^{-52}$\\
\hline
Error & $12.6\%$ & $15.7\%$ & $18.8\%$ & $21.5\%$ & $23.7\%$ & $25.2\%$ & $26.0\%$ \\
\hline \hline
$k$ & $14$ & $15$ & $16$ & $17$ & $18$ & $19$ & $20$ \\
\hline
$D_{2k}/\mathcal{N}_F$ & $2.64105405\cdot 10^{-55}$ & $9.29940331\cdot 10^{-59}$ & $3.65639217\cdot 10^{-62}$ & $1.60895318\cdot 10^{-65}$ & $7.92788909\cdot 10^{-69}$ & $4.36750842\cdot 10^{-72} $ & $2.68123444\cdot 10^{-75}$\\
\hline
Error & $26.2\%$ & $25.9\%$ & $25.4\%$ & $24.1\%$ & $24.3\%$ & $23.8\%$ & $23.5\%$ \\
\hline
\end{tabular}
\end{table*}

Under this assumption, the poor statistics for the computation of high order moments mentioned in the preceding paragraph is harmless. It also serves to show that the relationship between the quality of the reconstructed spectral functions and the relative errors of the moments is far from being a linear one. However, we do need to worry about the fact that the estimating functionals for the high order derivatives appearing in Eq.~(\ref{eq:25}) are generally not positive quantities. As such, there is the real possibility that, due to large cancellations between the positive and negative parts, the computed sequence $D_0, D_1, D_2, \ldots$ may not be a moment sequence.  In fact, taking into account the exponential instability of the matrices $\Delta_n$, one needs to worry about the limitations caused by the utilization of a finite Trotter index as well as about the inherent numerical limitations of the pseudo-spectral methods utilized. All these systematic errors are additional to the statistical errors. It appears then quite surprising that the present approach is extremely capable in this respect. The reader may use the results in Table~\ref{Tab:I} to verify that the stability matrix $\Delta_{20}$ is, indeed, positive definite (due to their structure, the stability matrices $\Delta_n$ for $n = 1, 2, \ldots 19$ are also positive definite). We mention that we have verified the positive definiteness of the matrices $\Delta_{20}$ (by performing a Cholesky decomposition\cite{Pre92}) not only for the final data, but also for the individual averages collected for each of the $100$ blocks in which the Monte Carlo simulation has been divided!  Why the algorithm is so capable in dealing with the sign problem is something that the author cannot explain at the present time. 

\begin{figure}[!tbp]
     \includegraphics[angle=270,width=8.5cm,clip=t]{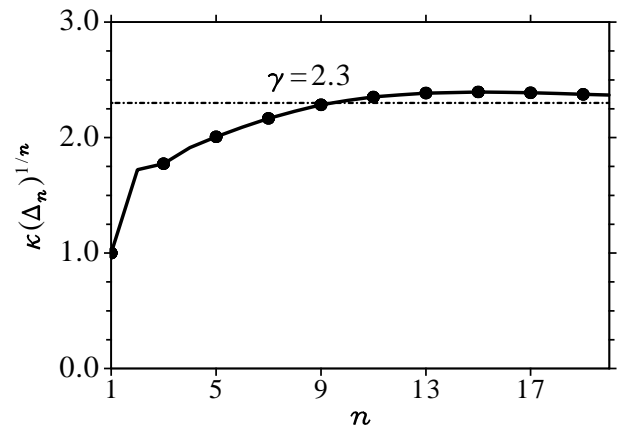}
   \caption[sqr] {\label{Fig:2} Asymptotic behavior of the quantities $\kappa(\Delta_n)^{1/n}$ for the flux-flux spectral function of the Eckart barrier. As for the free particle case, we notice that the matrices $\Delta_n$ become exponentially unstable. The condition number of the stability matrices increases roughly as $2.3^n$ for large $n$. }
\end{figure}

As demonstrated by the results in Table~\ref{Tab:I}, the relative statistical errors increase quickly with the order of the moments and reach a plateau at about $0.25-0.26$. This behavior seems to be caused by poor statistics, in a way that is perhaps similar with the previously discussed case,  where the spectral function is sampled directly. We have noticed that, for $k \geq 6$, the block averages fail to become independent and the Monte Carlo correlation times are comparable to the length of the simulation. It is therefore of certain interest to design alternative sampling techniques that would improve the statistics by use of suitable importance functions. However, the techniques should preserve the property of positive semi-definiteness of the stability matrices. Such a task appears formidable because the condition numbers of the stability matrices increase exponentially, according to the law $2.3^n$. This exponential instability is apparent from Fig.~\ref{Fig:2}.

\section{Summary and Conclusions}

We have provided an in-depth analysis of the problem of constructing estimators for the purpose of computing moments of spectral functions by path integral simulations. The estimators are constructed by formal differentiation of a certain estimating functional against the imaginary time. We have argued that the numerical differentiation can be more successfully implemented by means of pseudospectral methods, in a way that utilizes information from the entire interval $(-\beta \hbar / 2, \beta \hbar/2)$. The algorithmic detail that leads to robust numerical approximations is the fact that the action-like functionals and not the actual estimating functional are interpolated. The derivatives at the origin can be computed from the ensuing non-linear approximation in a fast and stable way by contour integration in the complex plane, with the help of  Lyness' method. 

We have improved upon the convergence results of Ref.~\onlinecite{Pre04-70} by stating Theorem~1, which provides the necessary and sufficient conditions for the existence of convergent reconstruction algorithms. The hypothesis of the new theorem is now based on  assertions that are verifiable solely from the computed data. In particular, we have pointed out that the existence of inverse algorithms is inherently related to the positive semi-definiteness of certain matrices, which, however, prove to be exponentially unstable. Although the main statements that lead to Theorem~1 are well-known results from the mathematical literature, we believe it is worth having them systematized in a single statement, for the benefit of the readers. 

One of the main assumptions made throughout the present work is that the potential function is infinitely differentiable. This assumption is not necessarily a limitation if one takes into account the existence of the partial averaging technique,\cite{Dol85} which replaces the action-like variables by smooth versions obtained by convolution with Gaussians of certain widths. Such convolutions are, of course, differentiable infinitely many times.\cite{Pre02f} For most practical applications, the value of the Gaussian width remains orders of magnitude larger than the resolution capabilities of the working precision. In other words, the interpolation technique utilized still ``sees'' a smooth functional even for the largest numbers of path variables that make the approximation convergent for all practical purposes. For these reasons, the partial averaging technique can be thought of as the natural setting for the implementation of the present approach.

Several questions related to the moment approach remain to be answered. The first one asks for an explanation of why the pseudo-spectral technique utilized leads to estimators that, upon largely inaccurate Monte Carlo integration, still produce a sequence of moments. By ``largely inaccurate,'' we mean that the statistical errors are orders of magnitude larger than the working precision required by the observed exponential instability of the matrices $\Delta_n$. Even the systematic errors introduced by the Chebyshev approximation are significantly larger than the required precision. A second question asks why the sign problem that should have ruined the Monte Carlo simulation is actually very mild. A third question is related to the development of sampling techniques, perhaps by means of suitable importance functions, that would alleviate the poor statistics associated with the computation of high-order moments, yet would preserve the positive semi-definiteness of the stability matrices. A final task calls for the development of actual reconstruction techniques of the spectral functions from their moments and for studies of the suitability of the various techniques for specific problems. 

\begin{acknowledgments} This work was supported in part by the National Science Foundation Grant Number CHE-0345280, the Director, Office of Science, Office of Basic Energy Sciences, Chemical Sciences, Geosciences, and Biosciences Division, U.S. Department of Energy under Contract Number DE AC03-65SF00098, and the U.S.-Israel Binational Science Foundation Award Number 2002170.
\end{acknowledgments}

\end{document}